\documentclass[preprint]{aastex}

\begin{document}

\title{Reconstruction of Stellar Orbits Close to Sagittarius A*:
Possibilities for Testing General Relativity}

\author{P. Chris Fragile and Grant J. Mathews}
\affil{University of Notre Dame, Center for Astrophysics, Notre Dame, IN 46556}
\email{pfragile@nd.edu}

\newfont{\ms}{cmsy10 at 12pt}

\begin{abstract}
We have reconstructed possible orbits for a collection of stars located 
within 0.5 arcsec of Sgr A$^*$.  These orbits are constrained 
by observed stellar positions and angular proper motions.  
The construction of such orbits serves as a baseline from which to search 
for possible deviations due to the unseen mass distribution in the central 
1000 AU of the Galaxy.  We also discuss the likelihood that some of 
these stars may eventually exhibit detectable relativistic effects, 
allowing for interesting
tests of general relativity around the $2.6\times10^6M_\odot$ central object.
\end{abstract}

\keywords{black hole physics --- celestial mechanics, stellar dynamics --- Galaxy: center --- Galaxy: kinematics and dynamics ---
                  gravitation --- relativity}

\section{Introduction}
The generally accepted idea that supermassive accreting black holes 
power the highly energetic phenomena in active galactic nuclei has motivated a 
great deal of effort to gather information about these extraordinary objects.  
Speculation about the presence of a black hole at the center of our own 
Galaxy has been ongoing for over 20 years [see Genzel et al. (1996) or 
Kormendy \& Richstone (1995) and references therein for recent summaries].  
Although the Galactic center appears to be nearly radio dormant, 
there is ample evidence for the presence of a supermassive black hole.  
[See, however, Tsiklauri \& Viollier (1998) and Munyaneza, 
Tsiklauri, \& Viollier (1998) for discussions of other possible interpretations.]

One of the most efficient ways to constrain the possible existence 
of a massive object in the Galactic core is to carefully observe the 
orbital motions of the stars and gas closest to the Galactic center.  
Furthermore, deviations from ideal orbits may provide a probe of 
the distribution of dark matter around the Galactic center.  Progress 
toward the accumulation of the relevant observational data has recently been made by 
Eckart et al. (1995), Genzel et al. (1996), Eckart \& Genzel (1997), 
Genzel et al. (1997), and Ghez et al. (1998) 
who have obtained high angular resolution K-band 
images of the Galaxy's central stellar cluster.

Here we report on an attempt to reconstruct ideal orbits for the innermost 8 
stars in the Ghez et al. (1998) survey using relativistic equations of motion.  
There is an obvious difficulty in determining orbital parameters for stars with 
periods much longer than the observational baseline, particularly when the 
uncertainties in the locations of the stars at each epoch are comparable to 
the apparent motion of the stars between observing epochs.  
Nonetheless, orbit reconstruction is possible in principle, since orbital 
mechanics is deterministic once the position and velocity vectors for the 
relevant masses are given at one instant of time.

The 8 stars studied here are the ones for which the possible general relativistic 
effects should be greatest.  The high velocities of these stars, and 
as we shall see, the possibility of significant periapse precession could 
provide a groundwork for interesting tests of general relativity as future 
observations of these stars are made.  We also 
discuss other possible detectable relativistic effects which could occur if the stars 
pass near the massive object.  Jaroszy\'nski (1998) has presented Monte 
Carlo simulations to demonstrate the feasibility of measuring such effects.  
Jaroszy\'nski (1999) and Salim \& Gould (1999) also demonstrate 
how accurate determinations of the orbit parameters can be used to 
constrain distance and mass estimates for the Galactic center.  
Here, we assess what can and has been learned from the currently available data.  
This work can also serve as a foundation for future, more-detailed probes 
of the mass distribution of the Galactic core.  It is important to initially
 compute the two-body orbits of these stars around the central massive object, 
without considering stellar interactions or mass distribution effects.  
These two-body orbits provide a well-defined dynamical model against which to 
compare the actual astrometry and velocities of the objects.  
Any discrepancy or lack thereof bears directly on the mass distribution around the 
central black hole and could serve as an indirect means to detect the 
dark matter distribution in the Galactic center.  For completeness, 
we also consider the case in which a few percent of the mass is distributed on 
scales comparable to the semi-major axes of the orbits being considered.  
As will be seen, there are distinctly different observational consequences in 
each case.

In this paper, we present a summary of the available published astrometric data 
in {\ms \symbol{'170}} 2.  Section 3 describes our orbital solution technique, 
and {\ms \symbol{'170}} 4 gives the constraints on the orbit parameters.  
In {\ms \symbol{'170}} 5 we discuss the possible relativistic and 
hydrodynamic effects these orbits may display.  In {\ms \symbol{'170}} 6 we 
discuss which observations would be most beneficial in future studies.

\section{Astrometric Data}
Genzel et al. (1997) (hereafter referred to as GEOE) have presented astrometric 
$K$-band maps of the central $3\times3$ arcsec$^2$ of the Galaxy's 
central star cluster for five epochs between 1992 and 1996.  
These images were taken using the 3.5-m New Technology Telescope of the 
European Southern Observatory.  Ghez et al. (1998) (hereafter referred to as GKMB) 
have presented $K$-band maps of the central $1\times1$ arcsec$^2$ 
for three epochs between 1995 and 1997.  Their data was taken using the 
W. M. Keck 10-m telescope.  Each group reported on the RA and DEC separation from 
Sgr A$^*$ as well as the angular proper motions for the stars.  Genzel et al. (2000) presents an updated tabular summary of the proper motions obtained to date.
Fig. 1 shows the combined data for the 8 stars studied here.  
The error bars show the uncertainty in the centroid positions for 
each data point.  The reported uncertainties were typically on the 
order of $0.008$ arcsec for GEOE and $0.002$ arcsec for GKMB.  
We followed the naming convention of GKMB, so the 8 stars studied here are 
given the labels S0-1 through S0-8.  We have ignored the first epoch of data from 
GEOE (1992.65) because of the difficulty they had in resolving any of the 
individual stars in the central-most region during that observation.  
Also, star S0-8 was not evident in the 1994.27 epoch of the GEOE data.  
Both groups used the results from Menten et al. (1997) to determine the position 
of Sgr A$^*$.  This means that the RA and DEC positions from both studies are subject 
to the inherent uncertainty of that work.  Although Menten et al. (1997) give 
an error estimate of $0.03$ arcsec, the GKMB group was able to link the 
infrared and radio reference frames and use 2 SiO maser sources 
to identify the position of Sgr A$^*$ to within $0.01$ arcsec.

\section{Fitting Method}
Because these stars have high velocities and may pass quite close to the supermassive object at the Galactic center, we choose to evolve the orbits using relativistic equations of motion for a test object 
of negligible mass orbiting a Schwarzschild or maximal Kerr black hole.  Given the extremely 
large mass of the central object ($2.6\pm0.2\times10^6M_\odot$), the approximation 
that the stars move in a fixed background metric is certainly acceptable.  
Jaroszy\'nski (1998) has studied possible orbits of these stars in a 
Kerr background and concluded that the effects of black hole angular momentum are 
probably negligible.  We also have made a study of orbits around a 
maximal Kerr black hole.
As we will describe below,
we found no detectable difference between the Kerr and
Schwarzschild dynamics. On 
the other hand, Munyaneza et al. (1998) have demonstrated 
that noticeable differences in orbit characteristics are possible if an extended mass 
distribution is present instead of a single compact object.  
However, GKMB have placed significant constraints on the possibilities of 
an extended mass distribution.  For this reason, we consider two separate cases: 
one in which all of the mass is contained in the central black hole and 
one in which roughly 5\% of the central mass is contained in an ideal 
($\gamma = 5/3)$ gas cloud in 
hydrostatic equilibrium, distributed on a scale equal to the best-fit 
semi-major axis of star S0-2.  The remaining 95\% of the mass is still in a central 
black hole.  The gravitational potentials for these two cases are plotted in Fig. 2.  
As more and better data become available, refinements to the mass distribution may 
be detectable.  
For illustration, the equations of motion 
in Schwarzschild coordinates with the origin fixed at 
Sgr A$^*$ are written (Weinberg 1972, Sec. 8.4 \& 11.1)
\begin{eqnarray}
{d^2r \over d\tau^2}&=&{-1 \over 2A(r)}{dA(r) \over dr}\left({dr \over d\tau}\right)^2 + {r \over A(r)}\left({d\theta \over d\tau}\right)^2 \nonumber \\ & + & {r \sin^2\theta \over A(r)} \left({d\phi \over d\tau}\right)^2 - {1 \over 2A(r)}\left({dB(r) \over dr}\right) \left({dt \over d\tau}\right)^2,
\end{eqnarray}

\begin{equation}
{d^2\theta \over d\tau^2}=-{2 \over r}{d\theta \over d\tau}{dr \over d\tau} + \sin\theta \cos\theta \left({d\phi \over d\tau}\right)^2,
\end{equation}

\begin{equation}
{d^2\phi \over d\tau^2}= -{2 \over r}{d\phi \over d\tau}{dr \over d\tau} - 2 \cot\theta\left({d\phi \over d\tau}\right)\left({d\theta \over d\tau}\right),
\end{equation}
and
\begin{equation}
{d^2t \over d\tau^2}=-{d\ln B(r) \over dr}\left({dt \over d\tau}\right)\left({dr \over d\tau}\right),
\end{equation}
where the Schwarzschild metric parameters inside the mass distribution are
\begin{equation}
A(r)=\left[1-{2\mathcal{M}(r) \over r}\right]^{-1}
\end{equation}
and
\begin{equation}
B(r)=\exp \left\{-\int^\infty_r {2 \over {r'}^2}[\mathcal{M}(r')+4\pi {r'}^3 P(r')]\left[1-{2\mathcal{M}(r') \over r'}\right]^{-1} dr' \right\}
\end{equation}
where
\begin{equation}
\mathcal{M}(r)\equiv\int^r_04\pi r'^2 \rho(r') dr' ~.
\end{equation}
Outside the mass distribution,
\begin{equation}
B(r)=A^{-1}(r)=1-{2\mathcal{M}(R) \over r} ~.
\end{equation}
Here $r$, $\theta$, and $\phi$ have their usual meanings, $\tau$ is the proper time, $t$ is the coordinate time, $\mathcal{M}(R)=M$ is the mass of the central object, $P$ is the proper pressure, and $\rho$ is the proper total energy density.  Here and throughout we use the convention of Weinberg (1972), i.e. a time-like metric, a negative Riemann tensor, and a negative sign in the Einstein equation [cf. Misner, Thorne, \& Wheeler (1973), to convert to other conventions].  Also, we use geometrized units ($G=c=1$) except where otherwise noted.

Setting up the problem thus means that the parameters which specify a model orbit are $\{r_0,\theta_0,\phi_0,\dot{r}_0,\dot{\theta}_0,\dot{\phi}_0\}$, 
i.e.~the position and velocity components at some instant in time, $t_0$.  
We find these parameters using a least square minimization of the expression:
\begin{equation}
\chi^2=\sum_{i=1}^N{(\mathrm{RA}(t_i;r_0,\theta_0,\phi_0,\dot{r}_0,\dot{\theta}_0,\dot{\phi}_0)-\mathrm{RA}_i)^2+(\mathrm{DEC}(t_i;r_0,\theta_0,\phi_0,\dot{r}_0,\dot{\theta}_0,\dot{\phi}_0)-\mathrm{DEC}_i)^2 \over \sigma_i^2}
\end{equation}
where RA$_i$ and DEC$_i$ are the measured RA and DEC offsets at epoch $t_i$ and RA$(t_i;r_0,\theta_0,\phi_0,\dot{r}_0,\dot{\theta}_0,\dot{\phi}_0)$ and 
DEC$(t_i;r_0,\theta_0,\phi_0,\dot{r}_0,\dot{\theta}_0,\dot{\phi}_0)$ are the 
corresponding coordinate offsets for that same epoch, calculated using 
the model parameters.  The error bars in the measured RA and DEC offsets 
are generally unequal.
Therefore the best weighting factor for the goodness of fit is the radius of the error ellipse along the line 
formed by joining the observed position at an 
epoch with the modeled location of the star at that epoch.
Hence, we write the weighting factor for the orbit fits as
\begin{equation}
\sigma_i^2 = (\sigma_{RA_i}\cos \Phi_i)^2+(\sigma_{DEC_i}\sin \Phi_i)^2
\end{equation}
where $\Phi_i$ is the angle between the RA-axis and the line 
drawn between the observed  and model-orbit positions.  To convert from angular separations to physical distances, we adopt 8.0($\pm0.5$) kpc as the distance to the Galactic center (Reid, 1993).  The uncertainty in this distance is included in our error estimates.

It is important to note that the orbits thus determined are degenerate in the parameters $\theta_0$ and $\dot{\theta}_0$.  There is no way to determine from the available data whether or not a particular star is currently in front of or behind Sgr A*.  By choice, we have assumed that all of these stars currently lie in front.  This choice affects the orbit parameters given in the next section, but has no effect on the conclusions of this paper.  Refined spectroscopic studies will help to break this degeneracy.

\section{Computed Orbits}
Having determined the model orbits as described above, we then convert the 
orbit parameters to their more familiar form 
$\{a,e,i,\Omega,\omega,T\}$, where $a$ is the semi-major axis, 
$e$ is the eccentricity, $i$ is the inclination, $\Omega$ is the 
longitude of the ascending node, $\omega$ is the argument of periapse, 
and $T$ is the time of last periapse passage [cf. Taff (1985), for the 
analytic expressions needed to perform this conversion].  Table 1 gives the 
orbit parameters for stars S0-1 through S0-3.  Stars S0-4 through S0-8, 
though analyzed, are not listed in Table 1 because the orbit 
parameter $\theta_0$ for these stars is not constrained by the observations, 
i.e. all orbits with $0\le\theta_0<\pi$ are within $1\sigma$ of the optimum fit.  
For the convenience of the readers, we have included the period, $P$, 
for each of the orbits in Table 1.  The errors given in the table are the 
statistical $1\sigma$ uncertainties for each parameter.  
Star S0-1 is an interesting case because its ``best-fit'' orbit 
is actually an unbound, hyperbolic trajectory.  Fig. 2 shows the computed, optimum least $\chi^2$ orbits overlaid on the observed data.

\section{Relativistic and Hydrodynamic effects}
The most relevant relativistic effect in a proper motion study is the precession 
of periapse.  The angular advance of the periapse, evaluated numerically here, 
is given approximately by
\begin{equation}
|\Delta\omega|=6\pi{GM \over a(1-e^2)c^2}
\end{equation}
in units of radians per revolution [cf. Ohanian \& Ruffini (1994)].  
It is a simple matter to transform the angular advance of periapse in 
the plane of the orbit into the more easily measured angular shift of apoapse in 
the plane of the sky.  Table 2 gives the periapse precession and the angular shift 
of apoapse for stars S0-2 and S0-3, assuming no mass distribution.  
For both stars, values near the $1\sigma$ upper limits are in principle measurable 
with the current observational position accuracies.  
Fig. 4 illustrates possible precession effects over multiple revolutions of a model 
orbit for star S0-2.  For this illustration, we show orbits confined to the plane 
of the sky ($i=0$), although the effect looks similar for any inclination in 
the range, $-40^\circ\lesssim i\lesssim40^\circ$.  
The $\chi^2$ for this orbit differs from the ``best-fit'' orbit by only 0.4.  
For comparison, we have shown a Newtonian orbit for two point-masses 
({\it left panel}), as well as relativistic orbits with and without distributed 
mass ({\it right panel} and {\it middle panel}, respectively).  
As expected, there is no precession in the Newtonian, point-mass limit.  
For the relativistic cases, there is a very important difference in the 
resulting precession.  Relativistic periapse precession, by itself, causes an 
angular {\bf advance} of periapse ({\it middle panel}).  
With the orbital inclination confined to small values, as in the model orbit 
illustrated here, the apoapse shift is $0.001$ arcsec per revolution and the orbital 
period is 10 years.  Using the current measured angular separation 
accuracy of $\pm0.002$ arcsec, the apoapse shift in this case could 
in principle be detectable with a baseline of $\approx20$ years of observation.  
The periapse precession due to a possible 
mass distribution, on the other hand, normally results in an 
angular {\bf regression} ({\it right panel}).  Therefore, a measured advance 
of periapse for these stars is most likely a relativistic effect.  
Such a measurement would also place strict limits on any mass distribution.  
Clearly this is a difficult measurement and one must carefully exclude competing 
effects from other masses around the central black hole.  
Nevertheless, it is worth searching for, since even determining the 
sign of the precession will reveal a great deal.

We also considered the effect of black hole angular momentum on this orbit.  Any effect will be most pronounced when the angular momentum axis of the orbit is aligned (or anti-aligned) with the angular momentum axis of the black hole.  In this case, both were chosen to lie along the line of sight.  For the model orbit shown in Fig. 4, the effect of the black hole angular momentum amounts to a 0.0001 arcsec shift of the apoapse after 3 revolutions ($\approx 30$ years).  Thus it represents a 3\% effect on top of the relativistic periapse precession.  Since our $1 \sigma$ upper limit on the relativistic apoapse shift for S0-2 and S0-3 is 0.001 arcsec per revolution, the best-case $1 \sigma$ upper limit on the apoapse shift due to black hole angular momentum is 0.00003 arcsec per revolution.  Measuring this effect would require observations on a time scale of hundreds of years at current astrometric accuracies.  Hence, we conclude that the orbits of these stars can probably not be used to detect black hole angular momentum.

Another potentially measurable effect is the gravitational 
redshift ($\Delta\nu/\nu=-GM/c^2R$) of the light emitted from these stars as they 
proceed through their orbits.  Note, however, that the expected special relativistic red- and blueshifts for these
stars can be much larger that any anticipated 
gravitational redshift and must therefore
be precisely accounted for before any
gravitational redshift will be discernible.  Table 2 contains a summary of both of these effects.

If these stars are sufficiently massive, it is possible that gravitational 
radiation from their final inspiral and plunge 
may be detectable by space-based interferometry such as the proposed LISA
mission (P. Bender 1999, private communication).  
Nevertheless, the time
scale for angular momentum loss by gravitational radiation for the
orbits considered here is greater than a Hubble time
($\sim 10^{15}$ yr). The most likely cause of
the final plunge will be scattering by a 2-body interaction somewhere
along the orbit. 

This raises the issue as to whether the time scale
for these stars to suffer small angle deflections from other stars
around the black hole is less than or comparable to the time scale
necessary to carry out a measurement of the periapse precession.
The magnitude of the acceleration on star $i$
due to the other stars in the field
\begin{equation}
a = \sum_j {m_j \over r_{ij}^2}
\end{equation}
will be dominated at any time by the nearest stars.  For the observed stars,
the typical nearest neighbor is separated by about 0.2 arcsec on the sky or about 
0.008 pc at 8 kpc.  Allowing for  eight times as many
other (non K-luminous) stars to be present, the
typical separation might be 0.004 pc with an associated
 mean acceleration due to nearest neighbors of
$\langle a \rangle \sim 9 \times
10^{-9}$ m/sec.  Assuming the most pessimistic case
that there is always a 1 M$_\odot$ star at this closest
distance with the same direction, then the time scale for 
a star to be deflected by a minimally detectable  0.002 arcsec 
($\approx 16$ AU) shift is then 
\begin{equation}
t = \sqrt{2 \Delta x/ \langle a \rangle} \approx {\rm~700~yrs} ~.
\end{equation}
Also, it should be noted that due to the motion around the
black hole, these stars are probably moving 
with rapid relative velocities $v \sim 0.01~c$.
Therefore, the radius of gravitational influence 
$G M_\odot/\langle v^2 \rangle$ is quite small ($\sim 10^4$ km).
Hence, unless there are extremely close encounters, we do not 
expect the mutual interactions to
dominate over the relativistic periapse precession
of interest here.
On the other hand, any such deviation, if detected, would most likely be
distinguishable from periapse precession in both
direction and magnitude and would
serve as a means to detect the presence of unseen matter.

The possibility of gravitational lensing of stars near the the Galactic center has 
been discussed elsewhere (Alexander \& Sternberg 1998; Salim \& Gould 1999; 
Jaroszy\'nski 1999; and Capozziello \& Iovane 1999).  
It has been concluded that this effect is probably negligible (Jaroszy\'nski 1999).

As another point of possible 
interest we consider the hydrodynamic effects of these stars passing 
close to the central object.  For a star of mass $M_*\ll M_{BH}$, the Roche-lobe radius, $R$, is given by
\begin{equation}
{R \over r}=0.49\left({M_* \over M_{BH}}\right)^{1/3},
\end{equation}
where $r$ is the radial distance from the star to the central object and $M_{BH}$ 
is the mass of the central object (King \& Done 1993).  This effect 
is strongly dependent on the periapse separation of these stars.  It is 
reassuring that none of the optimum fits in the present study are 
passing so close that they might have experienced Roche-lobe overflow. 
Also, since it is now evident (Figer et al. 2000; Eckart, Ott, \& Genzel 1999) that these stars are 
probably OB stars and not extended K-giants,  Roche overflow seems unlikely.
 Nevertheless, 
within the $1\sigma$ uncertainties deduced here, it is at least 
possible that one or 
more of these stars might approach or temporarily exceed their Roche limit 
during the next periapse passage.  This might be interesting to watch for, 
and if it occurred, it might lead to a burst of activity in the Galactic center.  

As another possible observable effect, we consider whether any of these stars could pass close enough to the central black hole to experience hydrodynamic distortion 
due to tidal interactions and whether these distortions could lead to observable oscillations in surface temperature or luminosity.  For 
distortions from an equilibrium  spherical shape of the form
\begin{equation}
r = R + \epsilon Y_l^m(\phi,\theta) ~,
\end{equation}
 where $Y_l^m(\phi,\theta)$ are spherical harmonics,
both the oscillatory period and the excitation time scale 
for these Kelvin  modes (Lang 1999) should be of order
\begin{equation}
P = { 2 \pi \over [ 2 l(l-1) G M_* / (2 l + 1) R^3]^{1/2}} ~.
\end{equation}
We estimate
$ P \lesssim 3$ days, which is too short compared to the time of periapse passage for such oscillations to be excited. 

\section{Discussion}
Given the interesting physics that might be gleaned from extended observations of 
these orbits, it is useful to summarize what steps must be taken 
to reduce the large statistical uncertainties in the orbital parameters.  
Of course, a great deal is to be learned by continued astrometry.
This is ultimately the only way in which the periapse precessions and/or effects
of the mass distribution can be determined.  In addition, it is crucial
to obtain accurate radial line-of sight velocities as well as
the radial distance to Sgr A$^*$.
 Current studies have provided fairly accurate measurements of 
RA, DEC, $V_{RA}$, and $V_{DEC}$, however, these must be complemented
by K-band spectroscopy.  Eckart et al. (1999) and Figer et al. (2000) have reported 
on high resolution infrared spectroscopy in the vicinity of Sgr A$^*$.  
 The spectra are consistent with all of  the stars in this sample being OB stars.
The strongest feature observed from this region 
is a weak Br$\gamma$ emission line.
However, as of yet no individual redshifts for these stars have 
been identified. What is needed are long integrations with high
spatial resolution.  
Clearly, it should be a high priority to obtain such redshifts.  If all six 
coordinates could be measured to comparable precision  
(e.g. $\sigma_x=\sigma_y=\sigma_z=0.002$ arcsec $\approx20$ AU and 
$\sigma_{V_x}=\sigma_{V_y}=\sigma_{V_z}\approx100$ km s$^{-1}$) then the 
$1\sigma$ uncertainties in the orbit parameters for star S0-2 would reduce 
to about 25\% of the best-fit values derived in this work.  If the velocity 
uncertainties were reduced to 50 km s$^{-1}$ then the uncertainties in 
orbit parameters for S0-2 would be reduced to about 20\%.  

If the line-of-sight velocity 
for each star could be accurately obtained directly from spectroscopic studies  
then the change in radial velocity, 
$dV_z/dt=-GMz/(x^2+y^2+z^2)^{3/2}$, 
over time might be used to find 
a value for the final unknown parameter - the line-of-sight separation between the 
star and Sgr A$^*$ (labeled $z$).  Fig. 4 illustrates graphically the relation
 between $dV_z/dt$ and $z$ for each of the stars.  From this we see that, 
in order to measure $z$ to an accuracy of $20$ AU, $dV_z/dt$ must be measured 
to an accuracy of $\sim7.5$ km s$^{-1}$ yr$^{-1}$, in the case of the star S0-2.  
This would require, for instance, two measurements of $V_z$ at accuracies of 
50 km s$^{-1}$ separated by about 9.5 years.  This time scale for accurate 
orbit determination is similar to the time scale derived by Salim \& Gould (1999) 
to use these orbits to better constrain the distance to Sgr A$^*$.

\section{Conclusion}
We have explored families of possible orbits for the 8 known stars located 
within $0.5$ arcsec of the Galactic center.  Because line-of-sight 
velocities have not  yet been obtained and only positions through the 1998
observing epoch are available, orbital parameters could only be constrained for 
3 of the stars.  The ideal orbits of these stars could be much better constrained
from relatively short baseline studies of line-of-sight velocities, although
continued observations of angular positions and angular proper motions are crucial
 if relativistic or mass-distribution effects are to be identified.  
These orbits as they now stand have at least the potential to display some 
extraordinary properties, 
including periods of less than 10 years and very high eccentricities.  
The current range of orbit parameters also allow for the possibility of 
interesting close encounters with the $2.6\times10^6M_\odot$ object at the 
Galactic core.  

It is clear from Eckart et al. (1999) and Figer et al. (2000) that these stars are probably
blue, luminous, and very young OB stars most likely 
formed from material infalling
toward the black hole.  A likely explanation is tidal disruption of
infalling rich OB associations by the central black hole. 
 Hence, it is reasonable for
the stars to have the highly eccentric orbits deduced here.

The possibility of these eccentric orbits passing close to the black hole opens
up the possibility for detection of relativistic effects such as 
periapse precession and/or gravitational redshift.  At the same time, continued
observations may also reveal the presence and distribution 
of unseen matter at the Galactic core.  
Hydrodynamic effects may be evident as well, if the stars pass close enough
to be tidally distorted.  Also, the star currently passing closest to Sgr A$^*$ in 
the plane of the sky may not be in a bound orbit around the central mass.
This introduces  the intriguing questions as to whether other members of this association are unbound.
If so, then the inferred virial mass of the central object may need revision.

In light of the many interesting puzzles highlighted in this paper, it is 
imperative that priority be given to continued astrometric,
as well as spectroscopic, study of these stars.  Such studies
will provide valuable knowledge about the distribution and the 
dynamical evolution of mass 
within the central core of the Galaxy.

\begin{acknowledgements}
The authors wish to acknowledge useful discussion with A. Ghez.  
We would also like to thank an anonymous referee for numerous helpful comments, 
particularly in regards to the importance of mass distribution effects.  
This work was supported by the National Science Foundation under grant PHY-97-22086.
\end{acknowledgements}

\clearpage

\figcaption{The measured positions of stars S0-1 through S0-8 based on the data of Genzel et al. (1997) (1994.27, 1995.60, 1996.25, and 1996.43 epochs) and Ghez et al. (1998) (1995.44, 1996.49, and 1997.36 epochs). \label{fig1}}
\figcaption{Gravitational potential (in geometrized units) for the two mass distributions considered in this paper.  The lower curve is the potential for a single $2.6\times10^6 M_\odot$ black hole at the Galactic center.  The upper curve is the potential for a $2.47\times10^6 M_\odot$ black hole with $0.13\times10^6 M_\odot$ of matter in a hydrostatic gas cloud extending out to 1000 AU. \label{fig2}}
\figcaption{Best-fit orbits for S0-1 through S0-3.  For S0-1, the best fit is actually an unbound, hyperbolic trajectory. \label{fig3}}
\figcaption{Multi-revolution trace of a possible model orbit for star S0-2 confined to the plane of the sky ($i=0$).  In all 3 panels, the initial orbit parameters are the same and the star orbits in a clockwise direction.  {\it Left panel:} Newtonian orbit for point particles (no precession).  {\it Middle panel:} angular advance of periapse (apoapse shifts to the right) for ideal test particle orbit around Schwarzschild black hole (no mass distribution).  {\it Right panel:} angular regression of periapse (apoapse shifts to the left) due to the distribution of roughly 5\% of the central mass over a volume on the order of the size of the test orbit. \label{fig4}}
\figcaption{Instantaneous time derivative of the line-of-sight velocity as a function of the line-of-sight separation from the central massive object. \label{fig5}}

\clearpage

\begin{deluxetable}{cccccccc}
\footnotesize
\tablecaption{Best-fit Orbit Parameters \label{tbl-1}}
\tablewidth{0pt}
\tablehead{
\colhead{Star ID} & \colhead{$a$ ($10^3$ AU)} & \colhead{$P$ (yrs)} & \colhead{$e$} & \colhead{$T$ (year)} & \colhead{$i$ (deg)} & \colhead{$\Omega$ (deg)} & \colhead{$\omega$ (deg)}
}
\startdata
S0-1 & $\infty$ & $\infty$ & 1.8($\pm3.7$) & NA & 112($\pm6$) & 213($\pm30$) & 193($\pm327$)\\
S0-2 & 1(${+11 \atop -6}$) & 19(${+312 \atop -185}$) & 0.5(${+9.8 \atop -5.0}$) & 1981($\pm304$) & 106($\pm96$) & 269($\pm16$) & 312(${+2870 \atop -1360}$)\\
S0-3 & 3(${+183 \atop -175}$) & 80(${+8650 \atop -8260}$) & 0.5(${+59 \atop -56}$) & 1988($\pm1810$) & 96($\pm44$) & 335($\pm1220$) & 89($\pm3720$)\\
\enddata
\end{deluxetable}

\clearpage
% Note to editor: This table may need to be published using landscape layout.

\begin{deluxetable}{ccccccccc}
\footnotesize
\tablecaption{Predicted Relativistic Effects \label{tbl-2}}
\tablewidth{0pt}
\setlength{\tabcolsep}{0.02in}
\tablehead{
 & & & \colhead{Periapse} & \colhead{Apoapse} & \colhead{$[\Delta \nu /\nu]_{GR}$} & \colhead{$[\Delta \nu /\nu]_{GR}$} & \colhead{$|\Delta \nu /\nu|_{SR}$} & \colhead{$|\Delta \nu /\nu|_{SR}$} \\
\colhead{Star ID} & \colhead{Periapse} & \colhead{Apoapse} & \colhead{Precession$^{\mathrm{a}}$} & \colhead{Shift$^{\mathrm{b}}$} & \colhead{at Periapse$^{\mathrm{c}}$} & \colhead{at Apoapse$^{\mathrm{c}}$} & \colhead{at Periapse$^{\mathrm{d}}$} & \colhead{at Apoapse$^{\mathrm{d}}$}\\
  & \colhead{($10^2$ AU)} & \colhead{($10^2$ AU)} & \colhead{(deg/rev)} & \colhead{(mas/rev)} & ($\times 10^{-5}$) & ($\times 10^{-5}$) & ($\times 10^{-3}$) & ($\times 10^{-3}$)
}
\startdata
S0-2 & 5(${+301 \atop -179}$) & 15(${+515 \atop -305}$) & 0.04(${+0.42 \atop -0.25}$) & 0.03(${+1.27 \atop -0.74}$) & -5(${+202 \atop -341}$) & -2(${+35 \atop -59}$) & 6(${+41 \atop -24}$) & 2($\pm24$)\\
S0-3 & 13($\pm455$) & 38($\pm3,210$) & 0.01(${+1.01 \atop -0.96}$) & 0.01(${+1.28 \atop -1.22}$) & -2(${+62 \atop -65}$) & -1(${+55 \atop -58}$) & 0.1($\pm79.9$) & 0.02($\pm80.1$)\\
\enddata
\tablenotetext{a}{Periapse precession is the angular displacement of periapse in the orbit plane per revolution.}
\tablenotetext{b}{Apoapse shift is the apparent motion of apoapse on the sky as seen by an observer on Earth.}
\tablenotetext{c}{Gravitational Redshift.}
\tablenotetext{d}{Relativistic Doppler Shift.  The sign can not be determined from available data.}
\end{deluxetable}

\begin{figure}
\plotone{f1.epsi}
\end{figure}

\begin{figure}
\plotone{f2.epsi}
\end{figure}

\begin{figure}
\plotone{f3.epsi}
\end{figure}

\begin{figure}
\plotone{f4.epsi}
\end{figure}

\begin{figure}
\plotone{f5.epsi}
\end{figure}

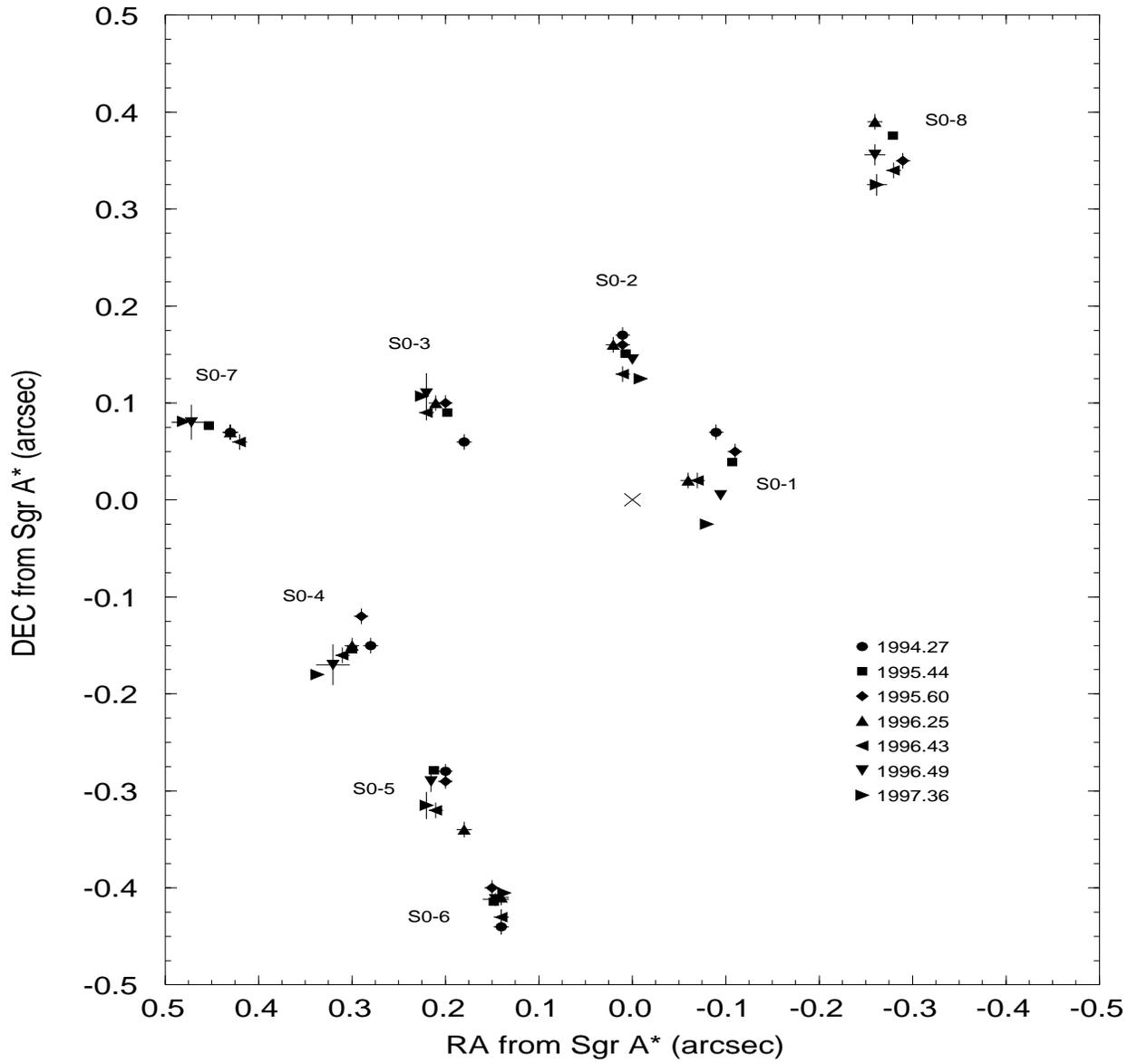
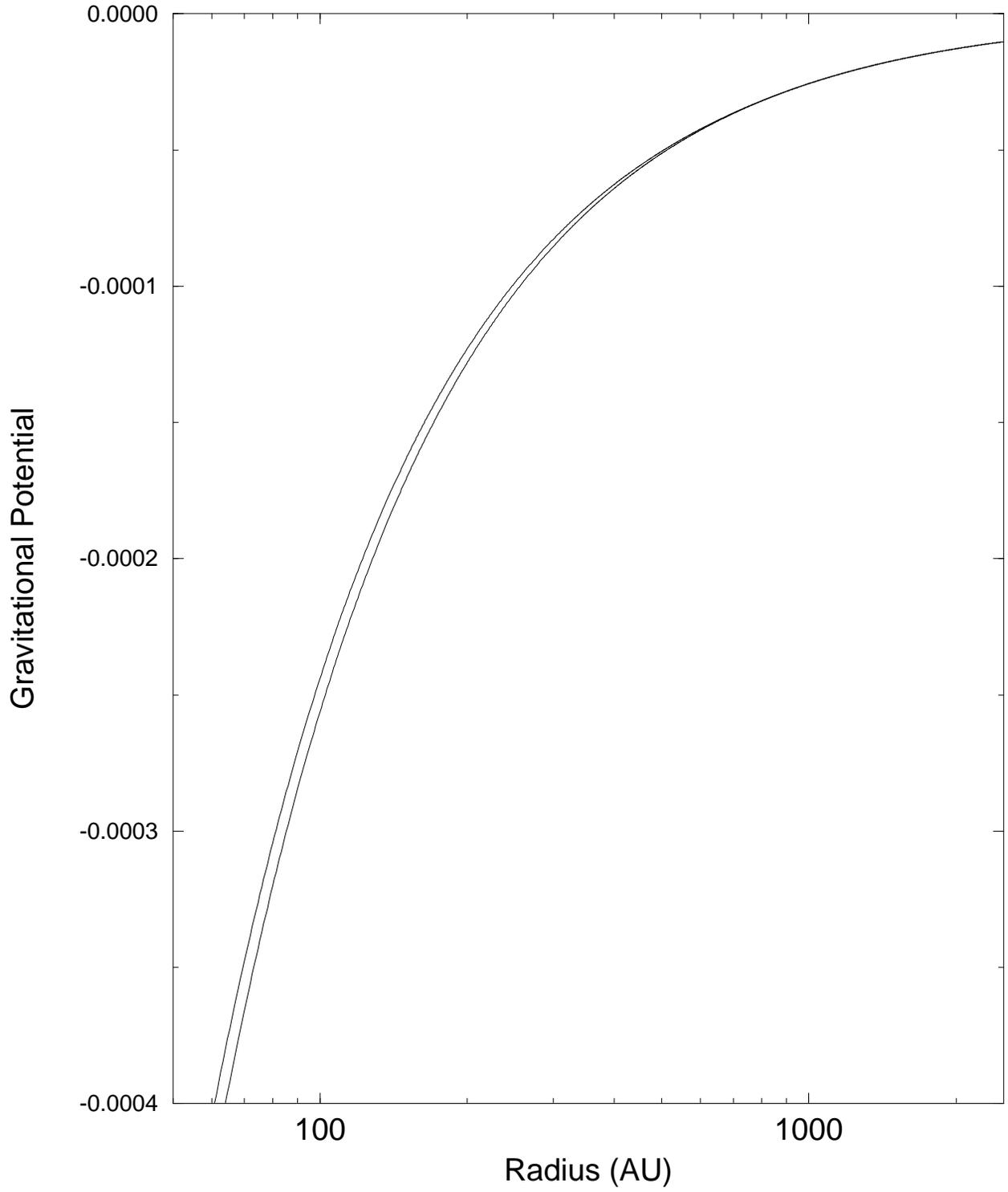
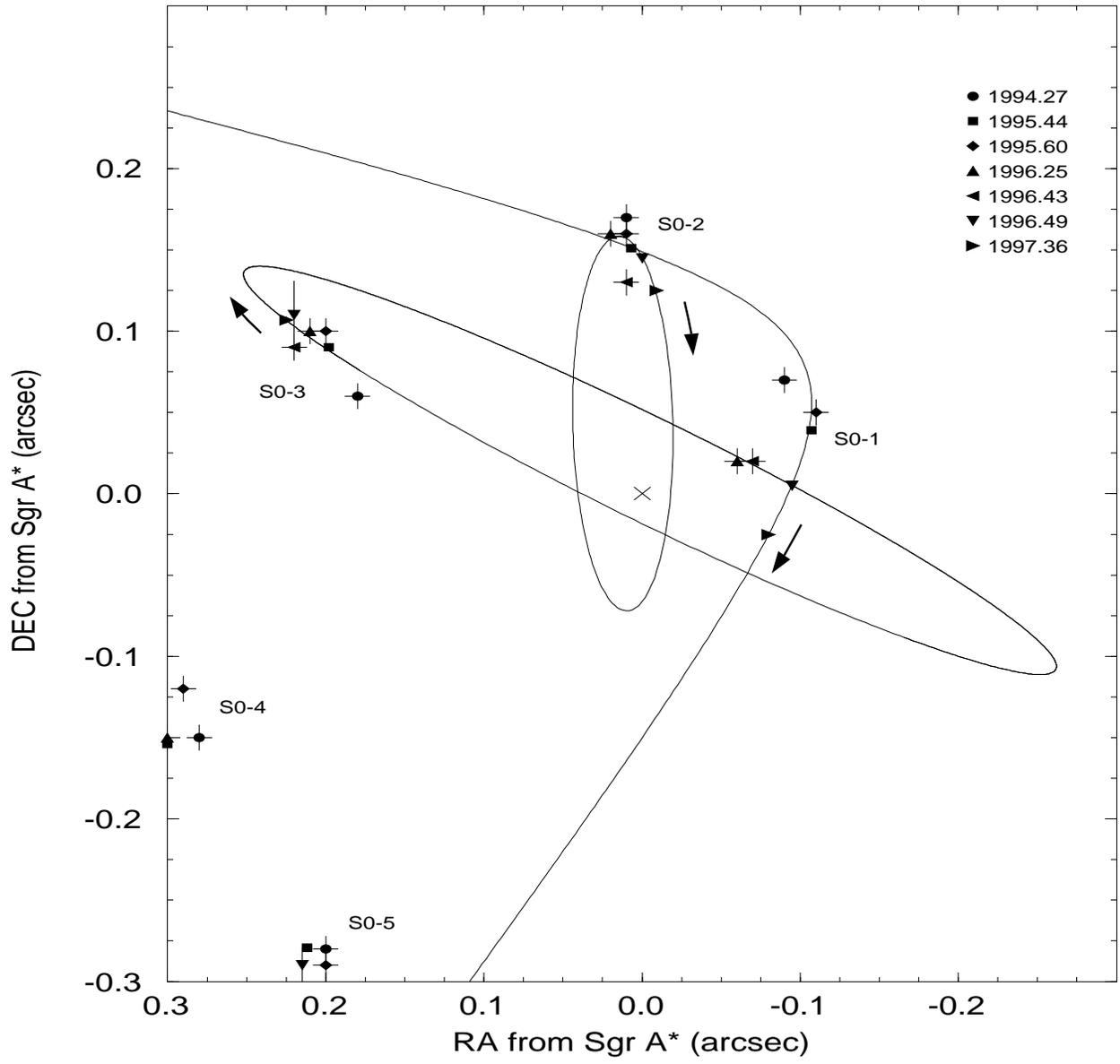
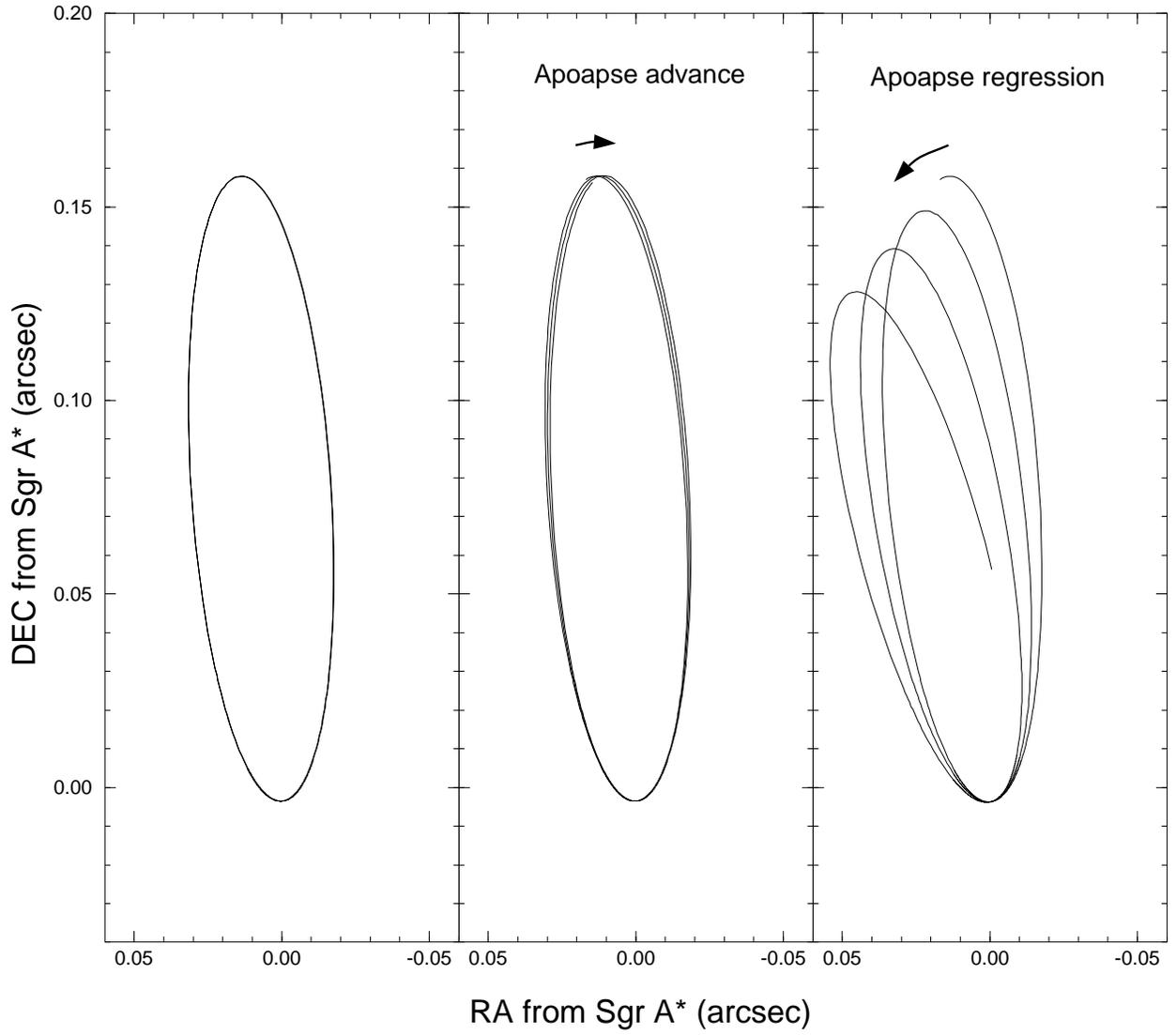
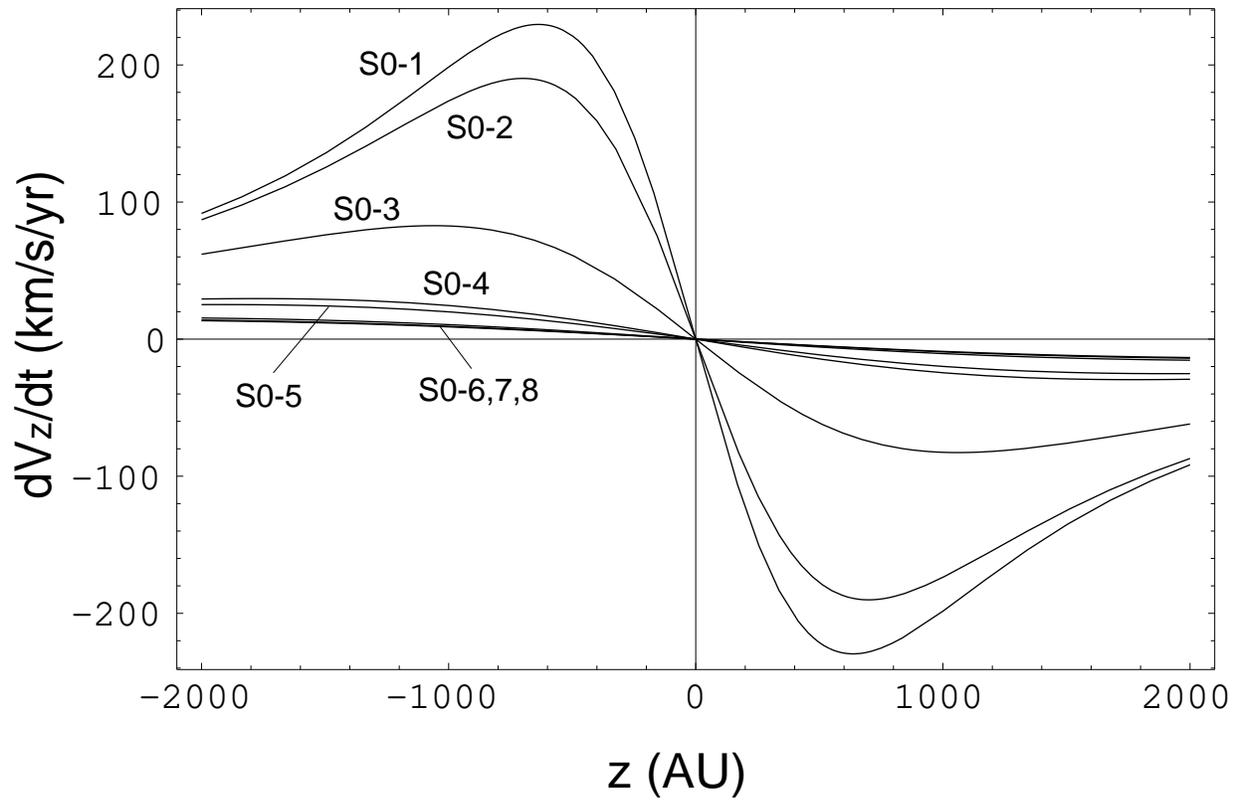
\end{document}